\documentstyle[emulateapj,apjfonts,psfig,amsmath,natbib]{article}
\def\ism{{\it ISM}}
\def\wind{{\it Wind}}

\begin{document}

\title{\large The Non-Relativistic Evolution of GRBs 980703 and
970508: Beaming-Independent Calorimetry}

\author{E.~Berger \& S.~R.~Kulkarni} 
\affil{Division of Physics, Mathematics, and Astronomy, 
California Institute of Technology 105-24, Pasadena, CA 91125}
\author{D.~A.~Frail}
\affil{National Radio Astronomy Observatory, P.~O.~Box O, Socorro,
NM 87801}

\begin{abstract} 
We use the Sedov-Taylor self-similar solution to model the radio
emission from the $\gamma$-ray bursts (GRBs) 980703 and 970508, when
the blastwave has decelerated to non-relativistic velocities.  This
approach allows us to infer the energy independent of jet collimation.
We find that for GRB\,980703 the kinetic energy at the time of the
transition to non-relativistic evolution, $t_{\rm NR}\approx 40$ d, is
$E_{\rm ST}\approx (1-6)\times 10^{51}$ erg.  For GRB\,970508 we find
$E_{\rm ST}\approx 3\times 10^{51}$ erg at $t_{\rm NR}\approx 100$ d,
nearly an order of magnitude higher than the energy derived in
\citet{fwk00}.  This is due primarily to revised cosmological
parameters and partly to the maximum likelihood fit we use here.
Taking into account radiative losses prior to $t_{\rm NR}$, the
inferred energies agree well with those derived from the early,
relativistic evolution of the afterglow.  Thus, the analysis presented
here provides a robust, geometry-independent confirmation that the
energy scale of cosmological GRBs is about $5\times 10^{51}$ erg, and
additionally shows that the central engine in these two bursts did not
produce a significant amount of energy in mildly relativistic ejecta
at late time.  Furthermore, a comparison to the prompt energy release
reveals a wide dispersion in the $\gamma$-ray efficiency,
strengthening our growing understanding that $E_\gamma$ is a not a
reliable proxy for the total energy.
\end{abstract}

\keywords{gamma-rays:bursts --- radiation mechanisms:nonthermal ---
shock waves}

\section{Introduction}
\label{sec:intro}

The two fundamental quantities in explosive phenomena are the kinetic
energy, $E_K$, and the mass of the explosion ejecta, $M_{\rm ej}$, or
equivalently the expansion velocity, $\beta\equiv v/c$, or Lorentz
factor, $\Gamma= (1-\beta^2)^{-1/2}$.  Together, these gross
parameters determine the appearance and evolution of the resulting
explosion.  Gamma-ray bursts (GRBs) are distinguished by a highly
relativistic initial velocity, $\Gamma_0\gtrsim 100$, as inferred from
their nonthermal prompt emission \citep{goo86,pac86}.  For the range
of $\gamma$-ray isotropic-equivalent energies observed in GRBs,
$E_{\rm\gamma,iso}\sim 10^{51}-10^{54}$ erg \citep{bfs01}, this
indicates $M_{\rm ej}\sim 10^{-5}-10^{-3}$ M$_\odot$, compared to
several M$_\odot$ in supernovae (SNe).

The true energy release of GRBs depends sensitively on the geometry of
the explosion.  For a collimated outflow (``jet'') with a half-opening
angle $\theta_j$, it is $E=f_bE_{\rm iso}$, where $f_b\equiv [1-{\rm
cos}(\theta_j)]$ is the beaming fraction; the true ejecta mass is also
a factor of $f_b$ lower.  Over the past several years there has been
growing evidence for such collimated outflows coming mainly from
achromatic breaks in the afterglow light curves
(e.g., \citealt{kdo+99,sgk+99}).  The epoch at which the break occurs,
$t_j$, corresponds to the time at which the ejecta bulk Lorentz factor
decreases below $\theta_j^{-1}$ \citep{rho99,sph99}.

In this context, several studies have shown that the beaming-corrected
energies of most GRBs, in both the prompt $\gamma$-rays and afterglow
phase, are of the order of $10^{51}$ erg
\citep{fks+01,pk02,bkf03,bfk03,yhs+03}.  The various analyses are
sensitive to the energy contained in ejecta with different velocities,
$\Gamma\gtrsim 100$ in the $\gamma$-rays, $\Gamma\gtrsim 10$ in the
early X-rays, and $\Gamma\gtrsim {\rm few}$ in the broad-band
afterglow.  However, {\it none} are capable of tracing the existence
and energy of non-relativistic ejecta.

\citet{fwk00} overcame this problem in the case of GRB\,970508 by
modeling the afterglow radio emission in the non-relativistic phase,
thus inferring $E_K\approx 5\times 10^{50}$ erg.  This analysis has
two significant advantages.  First and foremost it is independent of
jet collimation since the blastwave approaches spherical symmetry on
the same timescale that it becomes non-relativistic \citep{lw00}.
Second, this analysis relies on the simple and well-understood
Sedov-Taylor dynamics of spherical blastwaves, as opposed to the
hydrodynamics of spreading relativistic jets.  In addition, the peak
of the synchrotron spectrum on the relevant timescale lies in the
radio band where the afterglow is observable for several hundred days.

Two recent developments make similar analyses crucial.  We now
recognize that some GRBs are dominated by mildly relativistic ejecta
\citep{bkp+03}.  For example, for GRB\,030329 the kinetic energy
inferred from the afterglow emission, $E_K(\Gamma\sim {\rm few})
\approx 5\times 10^{50}$ erg \citep{bkp+03}, was an order of magnitude
higher than the $\gamma$-ray energy release \citep{pfk+03}.
Similarly, for GRB\,980425 $E_\gamma\approx 8\times 10^{47}$ erg
\citep{gvp+98,paa+00} was about $1\%$ of the relativistic kinetic
energy of the associated SN\,1998bw, $E_K\approx 10^{50}$ erg
\citep{kfw+98,lc99}.  This begs the question, is there even more
energy emerging from the engine, either at the time of the burst or
later on, at non-relativistic velocities?

Second, there is a growing interest in ``unification models'' for
GRBs, X-ray flashes (XRFs) and core-collapse SNe of type Ib/c, relying
primarily on energetics arguments.  For example, \citet{ldg03} argue
that GRBs and XRFs share an energy scale of $\sim 10^{49}$ erg, and
that all type Ib/c SNe give rise to GRBs or XRFs.  Both conclusions
result from significantly smaller values of $\theta_j$ compared to
those inferred in the past, such that the energy scale, $\propto
\theta_j^{2}$, is lower by a factor of $\sim 100$ and the true GRB
rate, $\propto\theta_j^{-2}$, matches locally the type Ib/c SN rate.
Given the important ramifications of the GRB energy scale for
progenitor scenarios we would like to independently address the
question: Is the energy scale of cosmic explosions $10^{49}$ erg,
implicating all type Ib/c SNe in the production of GRBs, or does it
cluster on $\sim 10^{51}$ erg?

The answer will also provide an independent confirmation of the jet
paradigm by comparison to the isotropic-equivalent energies.  This is
crucial since other explanations for the light curve breaks have been
suggested, including changes in the density of the circumburst medium,
a transition to a non-relativistic evolution on the timescale of a few
days (due to a high circumburst density), and changes in the energy
spectrum of the radiating electrons \citep{dl01,pan01,wl02}.

Here we address the possibility of significant contribution from
non-relativistic ejecta and robustly determine the energy scale of
GRBs independent of geometrical assumptions, using Very Large
Array\footnotemark\footnotetext{The VLA is operated by the National
Radio Astronomy Observatory, a facility of the National Science
Foundation operated under cooperative agreement by Associated
Universities, Inc.} radio observations of the afterglows of GRBs
970508 and 980703 in the non-relativistic phase.  We generally follow
the treatment of \citet{fwk00}, but unlike these authors we carry out
a full least-squares fit to the data.

\section{The Non-Relativistic Blastwave and Fireball Calorimetry}
\label{sec:method}

The dynamical evolution of an ultra-relativistic blastwave expanding
in a uniform medium (hereafter, \ism\,) is described in terms of its
Lorentz factor, $\Gamma=(17E_{\rm iso}/8\pi nm_pc^2r^3)^{1/2}$, where
$r$ is the radius of the blastwave and $n$ is the number density of
the circumburst medium \citep{bm76}.  This, along with the relation
for the observer time, which for the line of sight to the center of
the blastwave is $t\approx r/8\Gamma^2 c$ (e.g., \citealt{sar97}),
determines the evolution of the radius and Lorentz factor.  For a
spherical blastwave the expansion will eventually become
non-relativistic on a timescale\footnotemark\footnotetext{Here and
throughout the paper we use the notation $q=10^xq_x$.}, $t_{\rm
NR}\approx 65(E_{\rm iso,52}/n_0)^{1/3}$ d, determined by the
condition that the mass swept up by the blastwave, $M_{\rm sw}\approx
E_{\rm iso}/c^2$, where $M_{\rm sw}$.

An initially collimated outflow becomes non-relativistic at $t_{\rm
NR}\approx 40(E_{\rm iso,52}/n_0)^{1/4} t_{j,d}^{1/4}$ d \citep{lw00}.
Moreover, as the jet expands sideways (at $t\gtrsim t_j$) the outflow
approaches spherical symmetry on a timescale, $t_s\approx 150 (E_{\rm
iso,52}/n_0)^{1/4} t_{j,d}^{1/4}$ d, similar to $t_{\rm NR}$.  Thus,
regardless of the initial geometry of the outflow the non-relativistic
expansion is well-approximated as a spherical outflow.  We note that
this discussion can be generalized to a range of radial density
profiles.  Here, in addition to the \ism\ model, we focus on a density
profile, $\rho=Ar^{-2}$ (hereafter, \wind\,), appropriate for mass
loss with a constant rate, $\dot{M}_w$, and speed, $v_w$ \citep{cl00}.

Following the transition to non-relativistic expansion, the dynamical
evolution of the blastwave is described by the Sedov-Taylor
self-similar solution \citep{sed46,neu47,tay50}.  In this case the
radius of the shock is given by $r\propto (E_{\rm
ST}t^2/A)^{1/(5-s)}$, with $\rho=Ar^{-s}$.  Thus, in the \ism\ case
$r\propto (E_{\rm ST}t^2/nm_p)^{1/5}$, while in the \wind\ case
$r\propto (E_{\rm ST} t^2/A)^{1/3}$.  The constant of proportionality,
$\xi(\hat{\gamma})$, depends on the adiabatic index of the gas,
$\hat{\gamma}$, and is equal to 1.05 in the \ism\ case and 0.65 in the
\wind\ case for $\hat{\gamma}=13/9$.  The latter is appropriate for
pressure equilibrium between relativistic electrons and
non-relativistic protons\footnotemark\footnotetext{The relative
pressure between the protons and relativistic electrons depends on the
fraction of energy in relativistic electrons, $\epsilon_e$.  If this
fraction is low, the pressure may be dominated by the non-relativistic
protons in which case $\hat{\gamma}=5/3$.  As we show below,
$\epsilon_e$ for both GRB\,980703 and GRB\,970508 is in the range
$\sim 0.1$ to 0.5 and thus $\hat{\gamma}=13/9$ is applicable.}
\citep{fwk00}.  The circumburst material shocked by the blastwave is
confined downstream to a thin shell of width $r/\eta$, with
$\eta\approx 10$.

To calculate the synchrotron emission emerging from this shock-heated
material we make the usual assumptions.  First, the relativistic
electrons are assumed to obey a power-law distribution, $N(\gamma)
\propto \gamma^{-p}$ for $\gamma\ge\gamma_m$.  Second, the energy
densities in the magnetic field and electrons are assumed to be a
non-varying fraction ($\epsilon_B$ and $\epsilon_e$, respectively) of
the shock energy density.  Coupled with the synchrotron emissivity and
taking into account self-absorption, the flux received by an observer
at frequency $\nu$ and time $t$ is given by (e.g., \citealt{fwk00}):
\begin{equation}
F_\nu=F_0(t/t_0)^{\alpha_F}[(1+z)\nu]^{5/2}(1-e^{-\tau})
f_3(\nu/\nu_m) f_2^{-1}(\nu/\nu_m),
\label{eqn:ff}
\end{equation}
the optical depth is given by:
\begin{equation}
\tau_\nu=\tau_0(t/t_0)^{\alpha_\tau}[(1+z)\nu]^{-(p+4)/2}
f_2(\nu/\nu_m),
\label{eqn:tau}
\end{equation}
and the function
\begin{equation}
f_l(x)=\int_0^x F(y)y^{(p-l)/2}dy.
\label{eqn:f}
\end{equation}
Here, $\nu_m=\nu_0(t/t_0)^{\alpha_m}/(1+z)$ is the synchrotron peak
frequency corresponding to electrons with $\gamma=\gamma_m$, $F(y)$ is
given in e.g., \citet{rl79}, and the temporal indices $\alpha_F$,
$\alpha_\tau$ and $\alpha_m$ are determined by the density profile of
the circumburst medium.  In the \ism\ case $\alpha_F=11/10$,
$\alpha_\tau=1-3p/2$, and $\alpha_m=-3$, while in the \wind\ case
$\alpha_F=11/6$, $\alpha_\tau=-1-7p/6$, and $\alpha_m=-7/3$
\citep{wax04a}.  Equations~\ref{eqn:ff}--\ref{eqn:f} include the
appropriate redshift transformations to the rest-frame of the burst.

Based on the temporal scalings the synchrotron flux in the
optically-thin regime ($\nu\gg\nu_m,\nu_a$) evolves as $F_\nu\propto
t^{(21-15p)/10}$ (\ism\,) or $F_\nu \propto t^{(5-7p)/6}$ (\wind\,);
here the synchrotron self-absorption frequency, $\nu_a$, is defined by
the condition $\tau_\nu(\nu_a)=1$.  Thus, for $\nu\gg\nu_m,\nu_a$ the
transition to non-relativistic expansion is manifested as a steepening
of the light curves at $t_{\rm NR}$ if the outflow is spherical
\citep{spn98,cl00}, or a flattening if the outflow was initially
collimated \citep{sph99}.  Below, we use this behavior to estimate
$t_{\rm NR}$ for GRBs 980703 and 970508/

In \S\ref{sec:98} and \S\ref{sec:97} we use the temporal decay indices
and Equations~\ref{eqn:ff}--\ref{eqn:f} to carry out a least-squares
fit to the data at $t>t_{\rm NR}$ with the free parameters $F_0$,
$\tau_0$, $\nu_0$ and $p$.  These parameters are in turn used to
calculate the physical parameters of interest, namely $r$, $n_e$,
$\gamma_m$ and $B$; $n_e\approx (\eta/3)n$ is the shocked electron
density \citep{fwk00}.  Since only three spectral parameters are
available, this leaves the radius unconstrained and thus,
\begin{align}
B & = 11.7(p+2)^{-2}F_{0,-52}^{-2}(r_{17}/d_{28})^4 \,\,\, {\rm G},
\label{eqn:B} \\ 
\gamma_m & =6.7(p+2)F_{0,-52}\nu_{0,9}^{1/2}(r_{17}/d_{28})^{-2},
\label{eqn:gamma} \\  
n_e & =3.6\times 10^{10}c_n\eta_1 F_{0,-52}^3\nu_{0,9}^{(1-p)/2}
\tau_{0,32} r_{17}^{-1} (r_{17}/d_{28})^{-6} \,\,\, {\rm cm^{-3}},
\label{eqn:n} \\
c_n & =(1.67\times 10^3)^{-p}(5.4\times 10^2)^{(1-p)/2}(p+2)^2/(p-1).
\label{eqn:cn} 
\end{align}
In the \wind\ model, the density is appropriate at $r_{\rm ST}\equiv
r(t_{\rm NR})$, i.e., $\rho(r)=nm_p(r/r_{\rm NR})^{-2}$.

To determine the radius of the blastwave a further constraint is
needed.  We note that the energy contained in the electrons and
magnetic field cannot exceed the thermal energy of the Sedov-Taylor
blastwave, which accounts for about half of the total energy
\citep{fwk00}.  The energy in the electrons is given by $E_e=
[(p-1)/(p-2)]n_e\gamma_m m_ec^2 V$, while the energy in the magnetic
field is $E_B=B^2V/8\pi$; here $V=4\pi r^3/ \eta$ is the volume of the
synchrotron emitting shell.  Thus, using
Equations~\ref{eqn:B}--\ref{eqn:cn} and the condition $E_e+E_B\le
E_{\rm ST}/2$ we can constrain the range of allowed values of $r$.  In
the \ism\ model $E_{\rm ST}=nm_p(r/1.05)^5[t_{\rm NR}/(1+z)]^{-2}$,
while in the \wind\ model $E_{\rm ST}=A(r/0.65)^3[t_{\rm
NR}/(1+z)]^{-2}$.

With a constraint on the radius we can also ensure self-consistency by
calculating the velocity of the blastwave when it enters the
Sedov-Taylor phase, $v_{\rm ST}=2r(1+z)/5t_{\rm NR}$ (\ism\,) or
$v_{\rm ST}=2r(1+z)/3t_{\rm NR}$ (\wind\,).  We expect that roughly
$v\sim c$.  Finally, the isotropic-equivalent mass of the ejecta is
given by $M_{\rm ej}=4\pi nm_p r_{\rm ST}^3$ (\ism\,) or $4\pi Ar_{\rm
ST}$ (\wind\,).  The actual ejecta mass is reduced by a factor $f_b$
relative to this value.

\section{GRB\,980703}
\label{sec:98}

\begin{figure} 
\centerline{\psfig{file=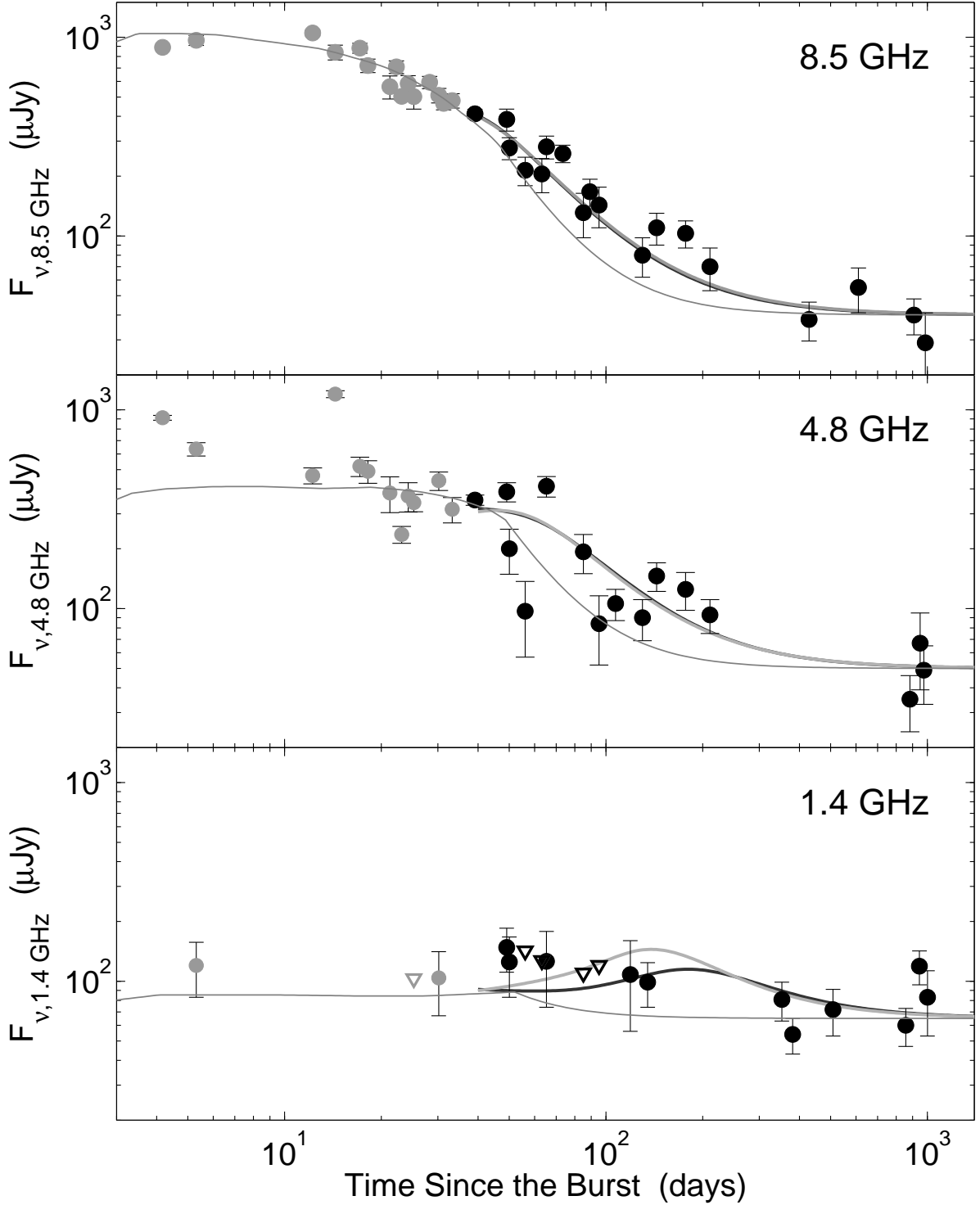,width=6.3in}}
\caption{Radio light curves of the afterglow of GRB\,980703 at 1.4, 4.9
and 8.5 GHz.  Only data at $t\ge t_{\rm NR}=40$ d (black circles) are
used in the fit.  The data exhibit a clear flattening relative to the
relativistic evolution of the afterglow (thin gray line;
\citealt{fyb+03}) in agreement with the expected change from $F_\nu
\propto t^{-p}$ (jet) to $F_\nu\propto t^{(21-15p)/10}$ (\ism\,) or
$F_\nu \propto t^{(5-7p)/6}$ (\wind\,) in the non-relativistic regime.
The best-fit light curves for the \ism\ (black) and \wind\ (gray)
models are indistinguishable.  The models include a contribution from
the host galaxy of 40, 50 and 65 $\mu$Jy at 8.5, 4.9 and 1.4 GHz,
respectively.
\label{fig:lc98}}
\end{figure}

In Figure~\ref{fig:lc98} we plot the radio light curves of GRB\,980703.
The data are taken from \citet{bkf01} and \citet{fyb+03}.  Two gross
changes in the light curves evolution are evident: a flattening at
$t\approx 40$ d at 4.9 and 8.5 GHz and a transition to a constant flux
density at late time.  The latter is due to radio emission from the
host galaxy of GRB\,980703 with flux densities at 1.4, 4.9 and 8.5 GHz
of 65, 50 and 40 $\mu$Jy, respectively \citep{bkf01}.  The flattening
at $t\approx 40$ d marks the transition to non-relativistic evolution
following a period of sideways expansion of the initially collimated
outflow (Figure~\ref{fig:lc98}).  A similar value, $t_{\rm NR}\approx
30-50$ d has been inferred by \citet{fyb+03} from tracking the
evolution of the blastwave Lorentz factor in the relativistic phase.
We therefore use here $t_{\rm NR}=40$ d.

We follow the method outlined in \S\ref{sec:method} using both the
\ism\ and \wind\ cases.  The results of both fits, shown in
Figure~\ref{fig:lc98}, are overall indistinguishable.  In what follows
we quote the results of the \ism\ model.  The best-fit parameters
($\chi^2_{\rm min}=123$ for 45 degrees of freedom) are:
$F_{0,-52}\approx 2.7$, $\tau_{0,32} \approx 80$, $\nu_{0,9} \approx
4.6$ and $p\approx 2.8$.  The relatively large values of $\chi^2_{\rm
min}$ is due primarily to fluctuations induced by interstellar
scintillation, particularly at 4.9 GHz.

\begin{figure} 
\centerline{\psfig{file=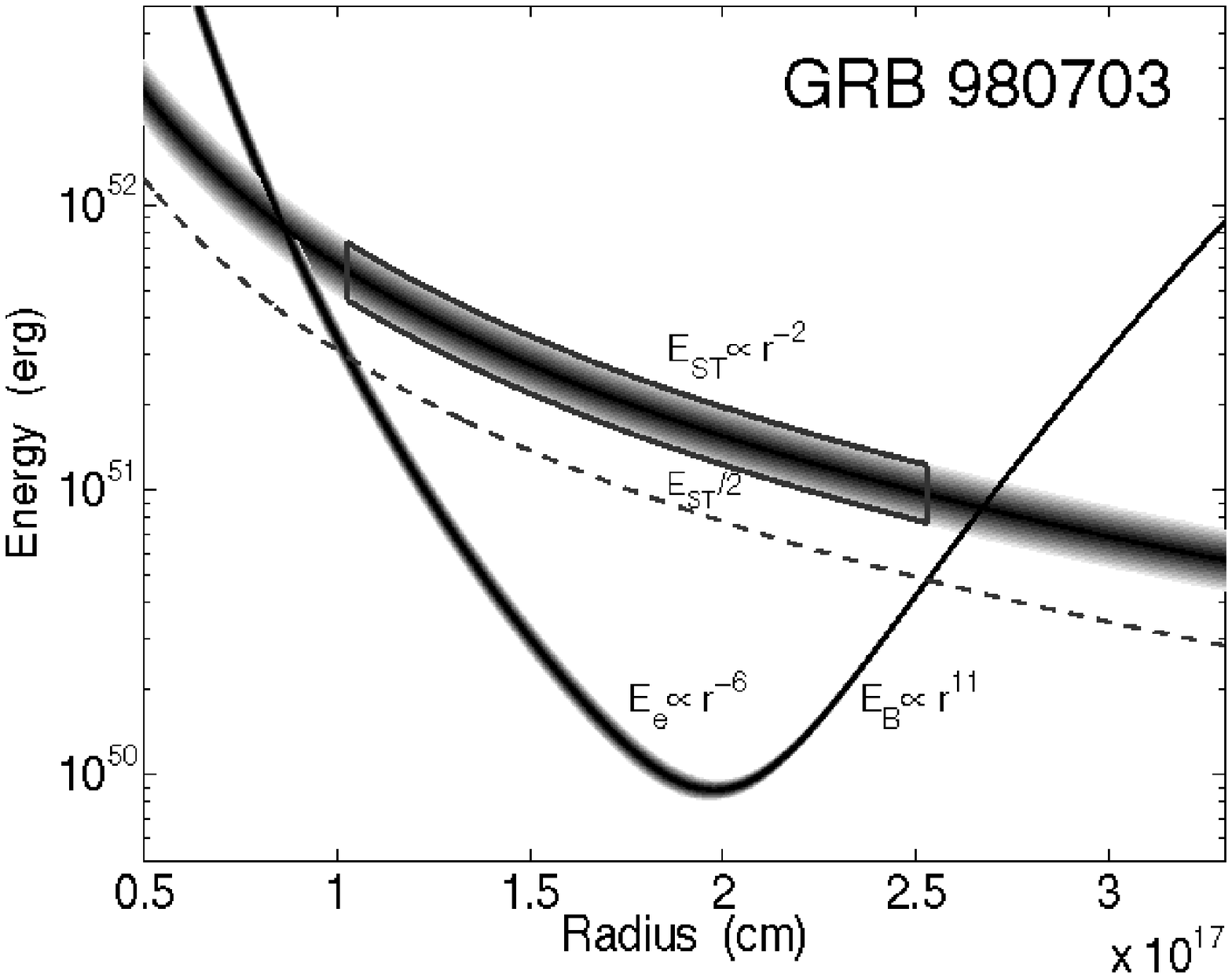,width=6.3in}}
\caption{Energies associated with the afterglow of GRB\,980703 in the
non-relativistic Sedov-Taylor phase as a function of the
(unconstrained) blastwave radius.  The thin curve is the sum of the
energy in relativistic electron ($E_e\propto r^{-6}$) and in the
magnetic fields ($E_B\propto r^{11}$).  Also plotted are the
Sedov-Taylor energy ($E_{\rm ST}\propto r^{-2}$) and the thermal
component, $E_{\rm ST}/2$.  The shading corresponds to an uncertainty
of $30\%$ in the value of the synchrotron frequency $\nu_{0}$ at
$t=t_{\rm NR}$.  The value of $E_{\rm ST}/2$ provides an additional
constraint, $E_e+E_B\le E_{\rm ST}/2$, which limits the range of
allowed radii in the solution (boxed region).
\label{fig:e98}}
\end{figure}

\begin{figure} 
\centerline{\psfig{file=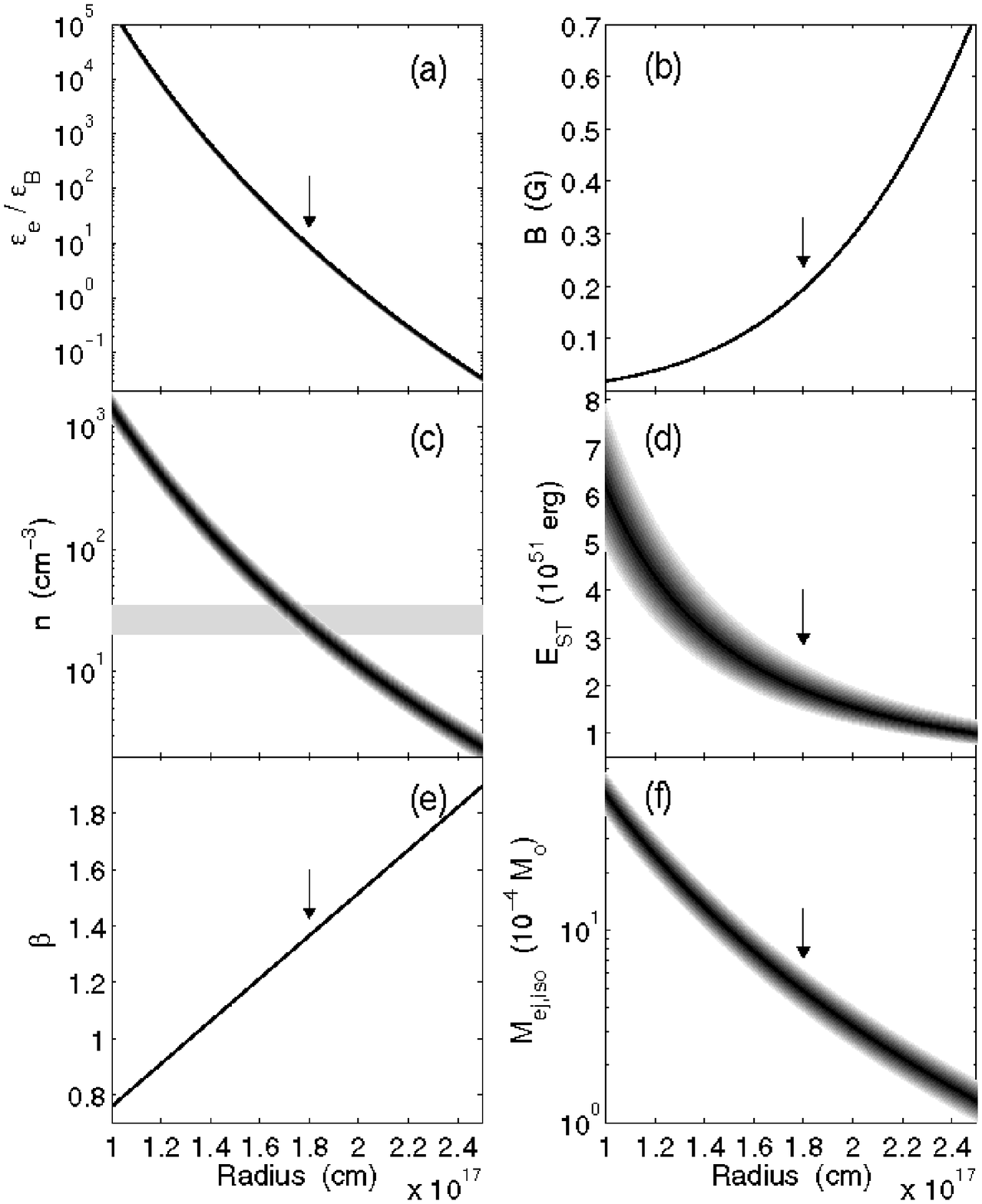,width=6.3in}}
\caption{Physical parameters of the Sedov-Taylor blastwave for
GRB\,980703 at $t_{\rm NR}=40$ d for the range of radii that obey the
constraint $E_e+E_B \le E_{\rm ST}/2$ (Figure~\ref{fig:e98}): (a) The
ratio of energy in the relativistic electrons to that in the magnetic
fields, (b) the magnetic field strength, (c) the density of the
circumburst medium, (d) the Sedov-Taylor energy, (e) the velocity of
the blastwave, and (f) the isotropic-equivalent mass of the ejecta
produced by the central engine and responsible for the afterglow
emission.  The light shaded region in (c) marks the range of densities
inferred from the relativistic evolution of the fireball, $n\approx
20-35$ cm$^{-3}$ \citep{fyb+03}.  With the additional constraint that
the density derived here conform to this value, we derive the values
of $\epsilon_e/\epsilon_B$, $B$, $E_{\rm ST}$, $\beta$, and $M_{\rm
ej}$ marked by arrows.
\label{fig:bn98}}
\end{figure}

Using $d_{28}=d_{L,28}/(1+z)^{1/2}=1.4$ ($z=0.966$, $H_0=71$ km
s$^{-1}$ Mpc$^{-1}$, $\Omega_m=0.27$ and $\Omega_\Lambda=0.73$), and
Equations~\ref{eqn:B}--\ref{eqn:cn} we find $B\approx 1.8\times
10^{-2}r_{17}^{4}$ G, $\gamma_m\approx 300r_{17}^{-2}$, and $n_e
\approx 4.9\times 10^3r_{17}^{-7}$ cm$^{-3}$.  From these parameters
we calculate $E_e\approx 3.4\times 10^{51}r_{17}^{-6}$ erg, $E_B
\approx 1.7\times 10^{46}r_{17}^{11}$ erg, and $E_{\rm ST}\approx
6.2\times 10^{51}r_{17}^{-2}$.  These results are plotted in
Figures~\ref{fig:e98}~and~\ref{fig:bn98}.

The range of blastwave radii allowed by the constraint $E_e+E_B
\lesssim E_{\rm ST}/2$ is $r_{17}\approx 1.05-2.5$, resulting in a
range of values for the Sedov-Taylor energy, $E_{\rm ST}\approx
(1-6)\times 10^{51}$ erg.  Given the strong dependence on radius, the
ratio of energy in the electrons to the energy in the magnetic field
ranges from $\epsilon_e/\epsilon_B\approx 0.03-9\times 10^4$, while
the specific values range from $\epsilon_e\approx 0.01-0.45$ and
$\epsilon_B\approx 5\times 10^{-6}-0.4$.  The circumburst density is
in the range $n\approx 8-3.5\times 10^3$ cm$^{-3}$, while the
blastwave velocity is $\beta_{\rm ST}\approx 0.8-1.9$.  Finally, the
isotropic-equivalent mass of the ejecta ranges from $(1-40)\times
10^{-4}$ M$_\odot$.

A comparison to the values derived by \citet{fyb+03} using modeling of
the afterglow emission in the relativistic phase is useful.  These
authors find $n\approx 30$ cm$^{-3}$, $\epsilon_e\approx 0.27$ and
$\epsilon_B\approx 2\times 10^{-3}$.  Using the same density in our
model (Figure~\ref{fig:bn98}), as required by the \ism\ density
profile, gives a radius $r_{17}\approx 1.75$ and hence
$\epsilon_e\approx 0.06$ and $\epsilon_B\approx 4\times 10^{-3}$, in
rough agreement; the energy is $E_{\rm ST}\approx 2\times 10^{51}$
erg. 

If we assume alternatively that the energy in relativistic electrons
and the magnetic field are in equipartition, we find $r_{17}\approx
2.05$.  In this case, $E_{\rm ST}\approx 1.5\times 10^{51}$ erg, $n
\approx 10$ cm$^{-3}$, $B\approx 0.3$ G, and $\epsilon_e=\epsilon_B
=0.03$.

\section{GRB\,970508} 
\label{sec:97} 

The non-relativistic evolution of GRB\,970508 was studied by
\citet{fwk00}.  These authors provide a rough model for the radio
emission beyond $t_{\rm NR}\approx 100$ d and argue that the
constraint $E_e+E_B\lesssim E_{\rm ST}/2$ requires the electron and
magnetic field energy to be in equipartition, $\epsilon_e=\epsilon_B
\approx 0.25$, with $E_{\rm ST}\approx 4.4\times 10^{50}$ erg.  Here
we perform a full least-squares fit, using $t_{\rm NR}=100$ d, and
find somewhat different results.  We use $t_{\rm NR}\approx 100$ d,
noting that for GRB\,970508 the outflow appears to be
weakly-collimated \citep{yhs+03}, and hence the transition is
manifested as a mild steepening of the light curves (see
\S\ref{sec:method}).

The best-fit parameters in the \ism\
model\footnotemark\footnotetext{We do not consider the \wind\ case
since in this model the observed decay rates at 4.9 and 8.5 GHz,
$F_\nu\propto t^{-1.2}$, require $p\approx 1.7$ and hence an infinite
energy.  This can be avoided by assuming a break in the electron
energy distribution at $\gamma_b>\gamma_m$ with a power law index
$q>2$, but we do not have the data required to constrain either
$\gamma_b$ or $q$.} ($\chi^2_{\rm min}=164$ for 58 degrees of freedom)
are: $F_{0,-52}\approx 38$, $\tau_{0,32}\approx 3.1\times 10^{-3}$,
$\nu_{0,9}\approx 3$ and $p\approx 2.17$.  The large value of
$\chi^2_{\rm min}$ is due primarily to interstellar scintillation.

In comparison, \citet{fwk00} use $F_{0,-52}\approx 41$, $\tau_{0,32}
\approx 5.3\times 10^{-3}$, $\nu_{0,9}\approx 9.5$, and they set
$p=2.2$; a solution with $\nu_{0,9}\approx 4.2$ is also advocated but
it is not used to derive the physical parameters of the blastwave.
The formal $\chi^2$ values for these solutions are $225$ and $254$,
respectively, somewhat worse than the solution found here.

As a result, we find that solutions away from equipartition are
allowed.  Adopting the cosmological parameters used by \citet{fwk00},
$H_0=70$ km s$^{-1}$ Mpc$^{-1}$, $\Omega_m=1$ and $\Omega_\Lambda= 0$,
we find $E_{\rm ST}\approx (6-11)\times 10^{50}$ erg, a factor of
about $20-100\%$ higher than the values inferred by these authors.

\begin{figure} 
\centerline{\psfig{file=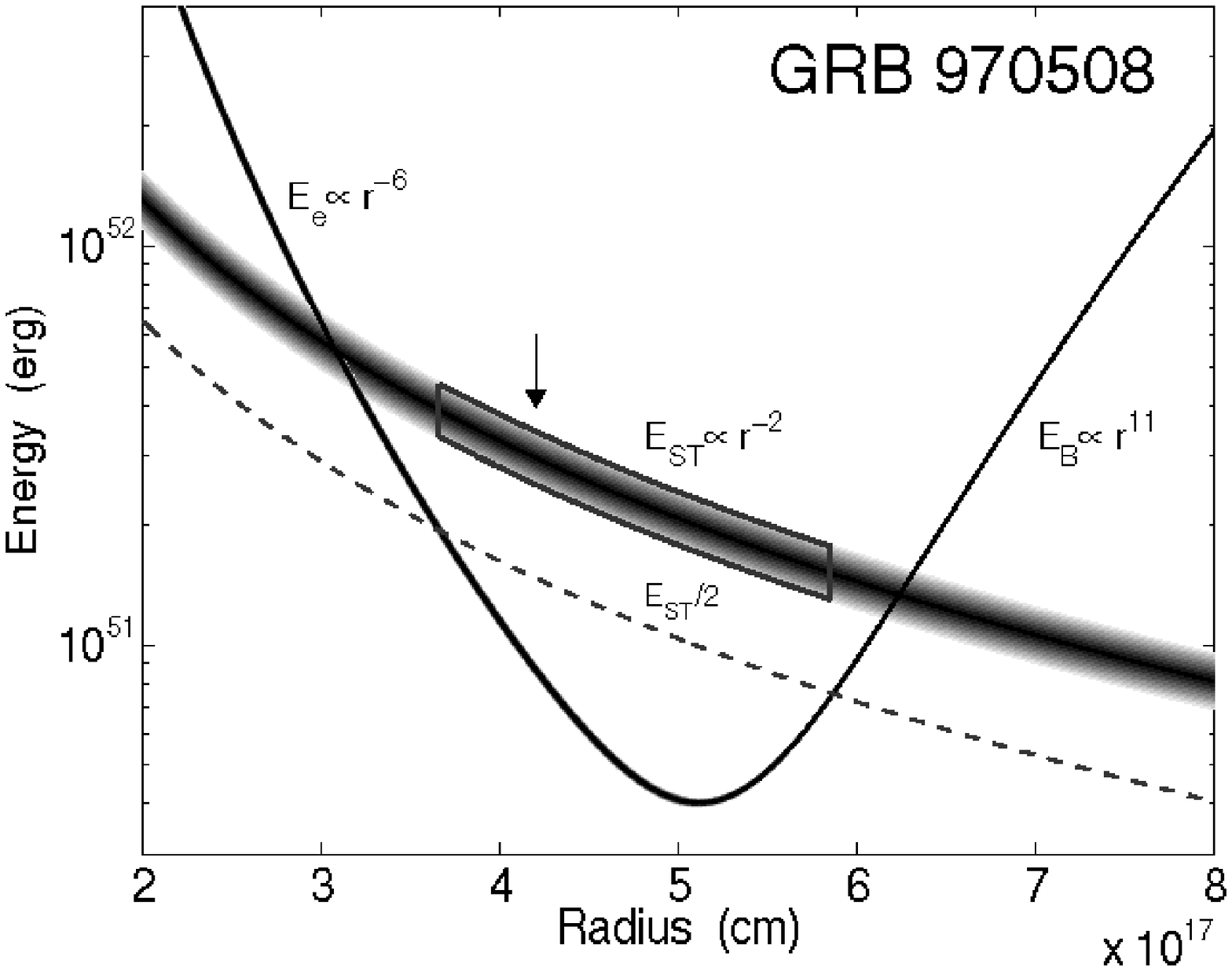,width=6.3in}}
\caption{Energies associated with the afterglow of GRB\,970508 in the
non-relativistic Sedov-Taylor phase as a function of the
(unconstrained) blastwave radius.  The thin curve is the sum of the
energy in relativistic electron ($E_e\propto r^{-6}$) and in the
magnetic fields ($E_B\propto r^{11}$).  Also plotted are the
Sedov-Taylor energy ($E_{\rm ST}\propto r^{-2}$) and the thermal
component, $E_{\rm ST}/2$.  The shading corresponds to an uncertainty
of $30\%$ in the value of the synchrotron frequency $\nu_{0}$ at
$t=t_{\rm NR}$.  The value of $E_{\rm ST}/2$ provides an additional
constraint, $E_e+E_B\le E_{\rm ST}/2$, which limits the range of
allowed radii in the solution (boxed region).  Finally, the arrow 
marks the most likely solution using the value of the cooling 
frequency as estimated from a combination of the radio and optical
data (\S\ref{sec:97}).  This additional parameter breaks the radius 
degeneracy, indicating $r\approx 4.2\times 10^{17}$ cm and $E_{\rm 
ST}\approx 3\times 10^{51}$ erg
\label{fig:e97}}
\end{figure}

\begin{figure} 
\centerline{\psfig{file=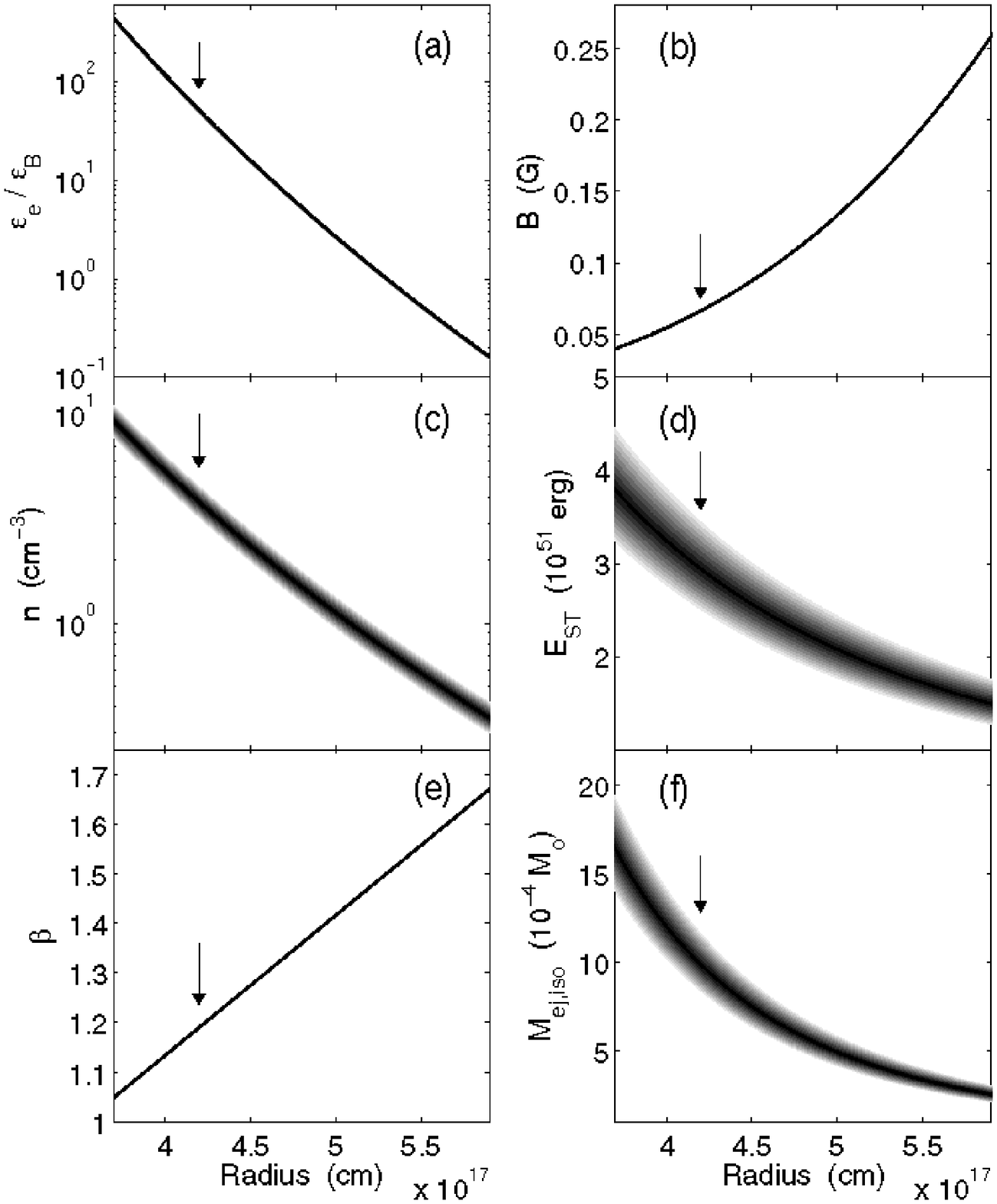,width=6.3in}}
\caption{Physical parameters of the Sedov-Taylor blastwave for
GRB\,970508 at $t_{\rm NR}=100$ d for the range of radii that obey the
constraint $E_e+E_B \le E_{\rm ST}/2$ (Figure~\ref{fig:e97}): (a) The
ratio of energy in the relativistic electrons to that in the magnetic
fields, (b) the magnetic field strength, (c) the density of the
circumburst medium, (d) the Sedov-Taylor energy, (e) the velocity of
the blastwave, and (f) the isotropic-equivalent mass of the ejecta
produced by the central engine and responsible for the afterglow
emission.  The arrows mark the most likely values using an estimate of 
the cooling frequency from a combination of the radio and optical   
data (\S\ref{sec:97} and Figure~\ref{fig:e97}).
\label{fig:bn97}}
\end{figure}

Using the currently-favored cosmology (\S\ref{sec:98}), we find
instead that the distance to the burst is higher by about $30\%$,
$d_{28}=1.21$ compared to $0.94$ \citep{fwk00}.  The change in
distance has a significant effect on the derived parameters since
$E_e\propto d^8$, $E_B\propto d^{-8}$ and $E_{\rm ST}\propto n\propto
d^6$.  Thus, we find that the constraint on $E_e+E_B$ indicates
$r_{17} \approx 3.7-5.9$ and therefore, $B\approx 0.04-0.25$ G,
$\gamma\approx 65-165$ and $n\approx 0.4-10$ cm$^{-3}$.  The
Sedov-Taylor energy is $E_{\rm ST}\approx (1.5-3.8)\times 10^{51}$
erg, while $\epsilon_e\approx 0.07-0.5$ and $\epsilon_B\approx
0.001-0.45$ (Figures~\ref{fig:e97}~and~\ref{fig:bn97}).  Assuming
equipartition, we find $r_{17}=5.3$, $E_{\rm ST}=1.8\times 10^{51}$
erg, and $\epsilon_e=\epsilon_b=0.11$.  The derived energy is about a
factor of four higher than the previous estimate \citep{fwk00}.

A comparison of our best-fit model with the flux of the afterglow in
the optical $R$-band at $t=110$ d, $F_{\nu,R}\approx 0.3$ $\mu$Jy
\citep{gcm+98}, indicates a break in the spectrum.  If we interpret
this break as due to the synchrotron cooling frequency, above which
the spectrum is given by $F_\nu\propto \nu^{-p/2}$, we find
$\nu_c\approx 6\times 10^{13}$ Hz.  Since $\nu_c=1.9\times 10^{10}
B^{-3}(t/110\,{\rm d})^{-2}$ Hz we infer $B\approx 0.073$ G and hence
$r_{17}=4.3$, $E_{\rm ST}=2.8\times 10^{51}$, $\epsilon_e=0.25$ and
$\epsilon_B=8\times 10^{-3}$.  These values are in rough agreement
with those inferred from modeling of the relativistic phase
\citep{pk02,yhs+03}, although our value of $\epsilon_B$ is somewhat
lower.

\section{Radiative Corrections}
\label{sec:rad} 

The energies derived in \S\ref{sec:98} and \S\ref{sec:97} are in fact
lower limits on the initial kinetic energy of the blastwave due to
synchrotron radiative losses.  These play a role primarily in the
fast-cooling regime ($\nu_c\ll \nu_m$), which dominates in the early
stages of the afterglow evolution (e.g., \citealt{spn98}).

\citet{yhs+03} estimate the time at which fast-cooling ends, $t_{\rm
cm}\approx 0.1$ and $1.4$ days after the burst for GRB\,970508 and
GRB\,980703, respectively.  Using these values, and our best estimate
of $\epsilon_e\approx 0.06$ (980703) and $\epsilon_e\approx 0.25$
(970508), we calculate the radiative corrections, $E\propto t^m$,
going back from $t_{\rm NR}$ to about 90 s after the burst.  Here
$m\approx -17\epsilon/12$, with $\epsilon=\epsilon_e/(1+
1.05\epsilon_e)$ for $t<t_{\rm cm}$ and it is quenched by a factor
$(\nu_m/\nu_c)^{(p-2)/2}<1$ at later times.  Thus, at low values of
$\epsilon_e$ the radiative losses are negligible.  The cutoff at 90 s
corresponds to the approximate deceleration time of the ejecta, $t_{\rm
dec}\approx 90(E_{52}/n_0\Gamma_2^8)^{1/3}$ s.

We find that approximately $50\%$ and $90\%$ of the energy was
radiated away before $t_{\rm NR}$ for GRBs 980703 and 970508,
respectively.  Thus, the initial kinetic energies are estimated to be
$4\times 10^{51}$ erg and $3\times 10^{52}$ erg, respectively.  The
corrections from $t_{\rm NR}$ back to $t_{\rm cm}$, $10\%$ for
GRB\,980703 and $70\%$ for GRB\,970508, indicate $E_K\approx 2\times
10^{51}$ and $9\times 10^{51}$ erg, respectively.  Both estimates of
the energy are in excellent agreement with those inferred from the
relativistic evolution of the fireball at $t_{\rm cm}$ \citep{yhs+03},
$E_K\approx 3\times 10^{51}$ erg (980703) and $E_K\approx 1.2\times
10^{52}$ erg (970508).

\section{Discussion and Conclusions}
\label{sec:conc}

Analysis of the synchrotron emission from a GRB blastwave in the
non-relativistic phase has the advantage that it is independent of
geometry and is described by the well-understood Sedov-Taylor
self-similar solution.  Using this approach to model the late-time
radio emission from GRBs 980703 ($t>40$ d) and 970508 ($t>100$ d) we
infer kinetic energies in the range $(1-6)\times 10^{51}$ erg and
$(1.5-4)\times 10^{51}$ erg, respectively.  Including the effect of
radiative losses starting at $t_{\rm dec}\sim 90$ s, we find that the
initial kinetic energies were about $4\times 10^{51}$ erg and $3\times
10^{52}$ erg, respectively.

The inferred kinetic energies confirm, independent of any assumptions
about the existence or opening angles of jets, that the energy scale
of GRBs is $\sim 5\times 10^{51}$ erg.  We therefore unambiguously
rule out the recent claim of \citet{ldg03} that the energy scale of
GRBs is of the order of $10^{49}$ erg.  Since the claimed low energies
were based on the apparent correlation between $E_{\rm \gamma,iso}$
and the energy at which the prompt emission spectrum peaks, $E_{\rm
peak}$ \citep{aft+02}, we conclude that this relation, and the prompt
emission in general, does not provide a reliable measure of the total
energy.  As a corollary, we rule out the narrow jet opening angles
used by \citet{ldg03}, $\theta_j\sim 0.1^{\rm o}$ and thus confirm
that the true GRB rate is significantly lower than the rate of type
Ib/c SNe \citep{bkf+03}.

Finally, the overall agreement between the energies derived here and
those inferred from modeling of the relativistic phase of the
afterglow indicates that the central engine in GRBs 980703 and 970508
did not produce a significant amount of energy in mildly relativistic
ejecta ($\Gamma\beta\gtrsim 2$) at late time, $t\sim t_{\rm NR}$.
However, a comparison to the beaming-corrected $\gamma$-ray energies
\citep{bfk03}, $E_\gamma \approx 1.1\times 10^{51}$ erg (GRB\,980703)
and $E_\gamma\sim 10^{51}$ erg (GRB\,970508) reveals that the
efficiency of the blastwave in producing $\gamma$-rays,
$\epsilon_\gamma$, varies considerably: $\sim 20\%$ for GRB\,980703,
but only $\sim 3\%$ for GRB\,970508.  The wide dispersion in
$\epsilon_\gamma$ strengthens the conclusion that $E_\gamma$ is not a
reliable tracer of the total energy \citep{bkp+03}.

The low value of $\epsilon_\gamma$ for GRB\,970508 may indicate an
injection of energy from mildly relativistic ejecta at early time.
Both the optical and X-ray light curves of this burst exhibited a
sharp increase in flux approximately 1 day after the burst, by a
factor of about 4 and $\gtrsim 2$, respectively \citep{paa+98,skz+98}.
The flux in these bands depends on energy as $F_\nu\propto
E^{(p+3)/4}$ and $\propto E^{(p+2)/4}$, respectively \citep{spn98}.
Thus, if we interpret the flux increase as due to injection of energy
from ejecta with $\Gamma\sim 5-10$ \citep{pmr98} we find an energy
increase of about a factor of three.  The analysis performed here
provides an estimate of the total energy following the injection and
thus $\epsilon_\gamma$ appears to be low.  The actual value of
$\epsilon_\gamma$ is thus $\sim 10\%$.

Although GRBs 980703 and 970508 are currently the only bursts with
sufficient radio data to warrant the full Sedov-Taylor analysis,
flattening of radio light curves at late time have been noted in
several other cases, most notably GRBs 980329, 991208, 000301C, 000418
and 000926 \citep{fmb+04}.  Interpreting the flattening as a
transition to non-relativistic expansion and using the expression for
the flux at $8.5$ GHz at the time of the transition, $F_\nu(t_{\rm
NR})\approx50[(1+z)/2]^{-1/2}\epsilon_{e,-1}\epsilon_{B,-1}^{3/4}
n_0^{3/4}E_{51}d_{28}^{-2}$ $\mu$Jy \citep{lw00}, we find the rough
results $n_0^{3/4}E_{51}\approx 6$ (980329), $\approx 4$ (991208),
$\approx 25$ (000301C), $\approx 6$ (000418), and $\approx 22$
(000926).  Thus, for typical densities, $\sim 1-10$ cm$^{-3}$
\citep{pk02,yhs+03}, the inferred kinetic energies are again of the
order of $10^{51}-10^{52}$ erg.

This leads to the following conclusions.  First, the energy scale of
cosmological bursts is about $5\times 10^{51}$ erg, at least three
orders of magnitude higher than the kinetic energies in fast ejecta
determined for local type Ib/c SNe from radio observations
\citep{bkc02,bkf+03}, and an order of magnitude higher relative to the
nearby ($d\approx 40$ Mpc) GRB\,980425 associated with SN\,1998bw
\citep{kfw+98,lc99,wax04b} and GRB\,031203 ($z=0.105$;
\citealt{pbc+04,skb+04}).  Second, as already noted in the case of
GRB\,030329 \citep{bkp+03}, there is a wide dispersion in the fraction
of energy in ultra-relativistic ejecta, such that the $\gamma$-rays
are a poor proxy for the total energy produced by the engine.

Thus, radio calorimetry is uniquely suited for addressing the relation
between various cosmic explosions.  So far, such studies reveal a
common energy scale in relativistic ejecta of about 5 foe (${\rm
foe}\equiv 10^{51}$ erg) for cosmological GRBs \citep{bkp+03}, about
0.1 foe for the low redshift bursts (980425, 031203), and $\lesssim
10^{-3}$ foe in fast ejecta for type Ib/c SNe.  The open question now
is whether we are beginning to trace a continuum in the energetics of
cosmic explosions, or whether the various classes truly represent
distinct physical mechanisms with different energy scales.
Fortunately, the best example to date of an object possibly bridging
the various populations, GRB\,030329, still shines brightly in the
radio a year after the burst.

\acknowledgements 

We thank Eli Waxman, Sarah Yost and Re'em Sari for valuable
discussions, and the referee, Roger Chevalier, for useful comments.
We acknowledge NSF and NASA grants for support.


\begin{thebibliography}{}

\bibitem[{Amati} {\it et al.}\ (2002)]{aft+02}
{Amati}, L. {\it et al.}\  2002, \aap, 390, 81.

\bibitem[{Berger}, {Kulkarni} \& {Chevalier}(2002)]{bkc02}
{Berger}, E., {Kulkarni}, S.~R., and {Chevalier}, R.~A. 2002, \apjl,
577, L5.

\bibitem[{Berger}, {Kulkarni} \& {Frail}(2001)]{bkf01}
{Berger}, E., {Kulkarni}, S.~R., and {Frail}, D.~A. 2001, \apj, 560,
652.

\bibitem[{Berger}, {Kulkarni} \& {Frail}(2003)]{bkf03}
{Berger}, E., {Kulkarni}, S.~R., and {Frail}, D.~A. 2003, ApJ, 590,
379.

\bibitem[{Berger} {\it et al.}\ (2003a)]{bkf+03}
{Berger}, E., {Kulkarni}, S.~R., {Frail}, D.~A., and {Soderberg},
A.~M. 2003a, \apj, 599, 408.

\bibitem[{Berger} {\it et al.}\ (2003b)]{bkp+03}
{Berger}, E. {\it et al.}\  2003b, \nat, 426, 154.

\bibitem[{Blandford} \& {McKee}(1976)]{bm76}
{Blandford}, R.~D. and {McKee}, C.~F. 1976, Physics of Fluids, 19,
1130.

\bibitem[{Bloom}, {Frail} \& {Kulkarni}(2003)]{bfk03}
{Bloom}, J.~S., {Frail}, D.~A., and {Kulkarni}, S.~R. 2003, \apj, 594,
674.

\bibitem[{Bloom}, {Frail} \& {Sari}(2001)]{bfs01}
{Bloom}, J.~S., {Frail}, D.~A., and {Sari}, R. 2001, \aj, 121, 2879.

\bibitem[{Chevalier} \& {Li}(2000)]{cl00}
{Chevalier}, R.~A. and {Li}, Z. 2000, ApJ, 536, 195.

\bibitem[{Dai} \& {Lu}(2001)]{dl01}
{Dai}, Z.~G. and {Lu}, T. 2001, \aap, 367, 501.

\bibitem[{Frail} {\it et al.}\ (2001)]{fks+01}
{Frail}, D.~A. {\it et al.}\  2001, ApJ, 562, L55.

\bibitem[{Frail} {\it et al.}\ (2004)]{fmb+04}
{Frail}, D.~A., {Metzger}, B.~D., {Berger}, E., {Kulkarni}, S.~R., and
{Yost}, S.~A. 2004, \apj, 600, 828.

\bibitem[{Frail}, {Waxman} \& {Kulkarni}(2000)]{fwk00}
{Frail}, D.~A., {Waxman}, E., and {Kulkarni}, S.~R. 2000, \apj, 537,
191.

\bibitem[{Frail} {\it et al.}\ (2003)]{fyb+03}
{Frail}, D.~A. {\it et al.}\  2003, \apj, 590, 992.

\bibitem[{Galama} {\it et al.}\ (1998)]{gvp+98}
{Galama}, T.~J. {\it et al.}\  1998, Nature, 395, 670.

\bibitem[{Garcia} {\it et al.}\ (1998)]{gcm+98}
{Garcia}, M.~R. {\it et al.}\  1998, \apjl, 500, L105+.

\bibitem[{Goodman}(1986)]{goo86}
{Goodman}, J. 1986, ApJ, 308, L47.

\bibitem[{Kulkarni} {\it et al.}\ (1999)]{kdo+99}
{Kulkarni}, S.~R. {\it et al.}\  1999, \nat, 398, 389.

\bibitem[{Kulkarni} {\it et al.}\ (1998)]{kfw+98}
{Kulkarni}, S.~R. {\it et al.}\  1998, \nat, 395, 663.

\bibitem[Lamb, Donaghy \& Graziani(2004)]{ldg03}
Lamb, D.~Q., Donaghy, T.~Q., and Graziani, C. 2004, {ApJ} (submitted),
astro-ph/0312634.

\bibitem[{Li} \& {Chevalier}(1999)]{lc99}
{Li}, Z. and {Chevalier}, R.~A. 1999, ApJ, 526, 716.

\bibitem[{Livio} \& {Waxman}(2000)]{lw00}
{Livio}, M. and {Waxman}, E. 2000, \apj, 538, 187.

\bibitem[{Paczynski}(1986)]{pac86}
{Paczynski}, B. 1986, ApJ, 308, L43.

\bibitem[{Panaitescu}(2001)]{pan01}
{Panaitescu}, A. 2001, \apj, 556, 1002.

\bibitem[{Panaitescu} \& {Kumar}(2002)]{pk02}
{Panaitescu}, A. and {Kumar}, P. 2002, ApJ, 571, 779.

\bibitem[{Panaitescu}, {Meszaros} \& {Rees}(1998)]{pmr98}
{Panaitescu}, A., {Meszaros}, P., and {Rees}, M.~J. 1998, \apj, 503,
314.

\bibitem[{Pian} {\it et al.}\ (2000)]{paa+00}
{Pian}, E. {\it et al.}\  2000, \apj, 536, 778.

\bibitem[{Piro} {\it et al.}\ (1998)]{paa+98}
{Piro}, L. {\it et al.}\  1998, \aap, 331, L41.

\bibitem[{Price} {\it et al.}\ (2003)]{pfk+03}
{Price}, P.~A. {\it et al.}\  2003, \nat, 423, 844.

\bibitem[{Prochaska} \& et~al.(2004)]{pbc+04}
{Prochaska}, J.~X. {\it  et~al.}\ 2004, {astro-ph/0402085}.

\bibitem[{Rhoads}(1999)]{rho99}
{Rhoads}, J.~E. 1999, \apj, 525, 737.

\bibitem[{Rybicki} \& {Lightman}(1979)]{rl79}
{Rybicki}, G.~B. and {Lightman}, A.~P. 1979, { {Radiative processes in
astrophysics}}, : New York, Wiley-Interscience, 1979.~393 p.).

\bibitem[{Sari}(1997)]{sar97}
{Sari}, R. 1997, \apjl, 489, L37+.

\bibitem[{Sari}, {Piran} \& {Halpern}(1999)]{sph99}
{Sari}, R., {Piran}, T., and {Halpern}, J.~P. 1999, ApJ, 519, L17.

\bibitem[{Sari}, {Piran} \& {Narayan}(1998)]{spn98}
{Sari}, R., {Piran}, T., and {Narayan}, R. 1998, ApJ, 497, L17+.

\bibitem[Sedov(1946)]{sed46}
Sedov, L.~I. 1946, Prikl.~Mat. i Mekh., 10, 241.

\bibitem[{Soderberg} \& et~al.(2004)]{skb+04}
{Soderberg}, A.~M. {\it  et~al.}\ 2004, {Submitted}.

\bibitem[{Sokolov} {\it et al.}\ (1998)]{skz+98}
{Sokolov}, V.~V., {Kopylov}, A.~I., {Zharikov}, S.~V., {Feroci}, M.,
  {Nicastro}, L., and {Palazzi}, E. 1998, \aap, 334, 117.

\bibitem[{Stanek} {\it et al.}\ (1999)]{sgk+99}
{Stanek}, K.~Z., {Garnavich}, P.~M., {Kaluzny}, J., {Pych}, W., and
{Thompson}, I. 1999, \apjl, 522, L39.

\bibitem[Taylor(1950)]{tay50}
Taylor, G.~I. 1950, Proc.~R.~Soc.~London A, 201, 159.

\bibitem[von Neumann(1947)]{neu47}
von Neumann, J. 1947, Los Alamos Sci.~Lab.~Tech.~Ser., 7.

\bibitem[{Waxman}(2004a)]{wax04a}
{Waxman}, E. 2004a, \apj, 602, 886.

\bibitem[{Waxman}(2004b)]{wax04b}
{Waxman}, E. 2004b, \apj, 605, L97.

\bibitem[{Wei} \& {Lu}(2002)]{wl02}
{Wei}, D.~M. and {Lu}, T. 2002, \mnras, 332, 994.

\bibitem[{Yost} {\it et al.}\ (2003)]{yhs+03}
{Yost}, S.~A., {Harrison}, F.~A., {Sari}, R., and {Frail}, D.~A. 2003,
\apj, 597, 459.

\end{thebibliography}

\clearpage
\begin{deluxetable}{lll}
\tabcolsep0.1in\footnotesize
\tablecolumns{3}
\tabcolsep0.2in\footnotesize
\tablewidth{0pc}
\tablecaption{Physical Parameters of GRBs 980703 and 970508
\label{tab:params}} 
\tablehead {
\colhead {Parameter}     &
\colhead {GRB\,980703}   &
\colhead {GRB\,970508} 
}
\startdata
$r$ ($10^{17}$ cm) & $1.05-2.5$ & $3.7-5.9$ \\ 
$B$ (G) & $0.02-0.7$ & $0.04-0.25$ \\ 
$\gamma$ & $8-270$ & $65-165$ \\ 
$n$ (cm$^{-3}$)	& $8-3.5\times 10^3$ & $0.4-10$ \\ 
$\epsilon_e$ & $0.01-0.45$ & $0.07-0.5$ \\ 
$\epsilon_B$ & $5\times 10^{-6}-0.4$ & $1\times 10^{-3}-0.45$ \\ 
$M_{\rm ej,iso}$ ($10^{-4}$ M$_\odot$) & $1-40$ & $2-18$ \\
$E_{\rm ST}$ ($10^{50}$ erg) & $9-56$ & $15-38$ \\ 
$E_{K}(t_{\rm dec})$ ($10^{51}$ erg) & $4$ & $30$
\enddata
\tablecomments{Physical parameters of GRBs 980703 and 970508 derived
from the non-relativistic evolution of their blastwaves.  The range of
allowed radii, and hence physical parameters, is determined by the
condition $(E_e+E_B)\le E_{\rm ST}/2$.  The last entry in the table,
$E_K(t_{\rm dec})$, is the total kinetic energy at the deceleration
time, $t_{\rm dec}\approx 90$ s, including synchrotron radiative
losses (\S\ref{sec:rad}).}
\end{deluxetable}

\clearpage

\clearpage

\clearpage

\clearpage
\begin{figure} 
\centerline{\psfig{file=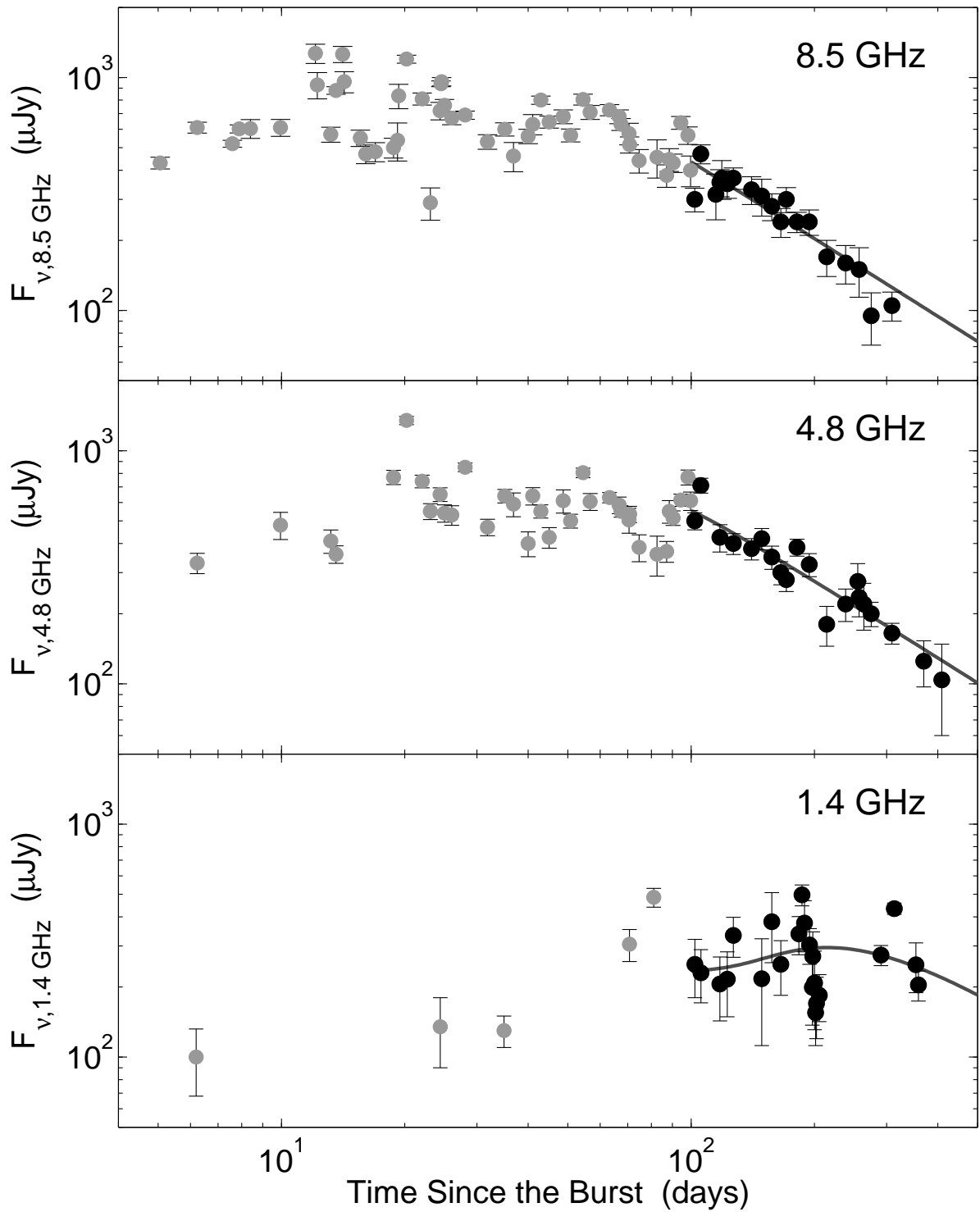,width=6.3in}}
\caption{Radio light curves of the afterglow of GRB\,970508 at 1.4, 4.9
and 8.5 GHz.  Only data at $t\ge t_{\rm NR}=100$ d (black circles) are
used in the fit.  The best-fit light curves for the \ism\ model are
shown (black); the \wind\ model can be ruled out since it requires
$p<2$.
\label{fig:lc97}}
\end{figure}

\clearpage

\clearpage

\end{document}